\documentstyle[12pt,harvard,epsf]{article}

\citationstyle{dcu}

\begin{document}
\title{Implementations of Quantum Logic: Fundamental and Experimental Limits}
\author{\em S. Bose, P.L. Knight, M. Murao, M.B. Plenio and V. Vedral\\ 
Optics Section, The Blackett Laboratory, \\Imperial 
College, London, SW7 2BZ, UK}


\maketitle
\begin{abstract}
Quantum information processing rests on our ability to manipulate
quantum superpositions through coherent unitary transformations. In
reality the quantum information processor (a linear ion trap, or
cavity qed implementation for example) exists in a dissipative
environment. Dephasing, and other technical sources of noise, as well
as more fundamental sources of dissipation severely restrict quantum
processing capabilities. The strength of the coherent coupling needed
to implement quantum logic is not always independent of dissipation.
The limitations these dissipative influences present will be described
and the need for efficient error correction noted. Even if long and
involved quantum computations turn out to be hard to realize, one can
perform interesting manipulations of entanglement involving only a few
gates and qubits, of which we give examples. Quantum communication
also involves manipulations of entanglement which are simpler to
implement than elaborate computations.  We briefly analyse the notion
of the capacity of a quantum communication channel.
\end{abstract}

\section{Introduction}
Since Shor's discovery \cite{PWS94,Ekert1} of an algorithm that allows
the factorisation of a large number by a quantum computer in
polynomial time instead of an exponential time as in classical
computing, interest in the practical realization of a quantum computer
has been much enhanced. Recent advances in the preparation and
manipulation of single ions as well as the engineering of pre-selected
cavity light fields suggests that quantum optics may well be that
field of physics promising the first experimental realization of a
quantum computer.

The realization of a quantum computer in a linear trap \cite{Ignacio}
has been regarded as very promising as it was thought that decoherence
could be suppressed sufficiently to preserve the superpositions
necessary for quantum computation. Indeed, a single quantum gate in
such an ion trap has been realized by Monroe et al
\cite{Monroe1}. Nevertheless, the error rate in this experiment was
too high to allow the realization of extended quantum networks. This
experiment was limited by technical difficulties and one aim of future
experiments is to reduce these to come closer to the fundamental
limits, such that at least small networks could be realized. However,
there remains the question whether overcoming technical problems will
be sufficient to realize practically useful computations such as
factorisation of big numbers on a quantum computer in a linear ion
trap. Here we address the problem of so called threshold accuracy in
quantum computation \cite{Knill1,Aharonov1}. This threshold implies
that arbitrarily complicated (long) quantum computations can be
performed once the error rate of a quantum gate can be pushed below a
certain threshold. We will discuss whether the required thresholds
\cite{Knill1,Aharonov1} can be achieved or if spontaneous emission
rules out this possibility (not to mention other error sources).  We
present a simple calculation to understand the order of magnitude of
these thresholds and then calculate the spontaneous emission rate in
one quantum gate. Even if long and involved quantum computations turn
out to be hard to realize, one can perform some interesting
manipulations of entanglement involving only a few gates and qubits,
of which we give some examples. Quantum communication also involves
manipulations of entanglement which are simpler to implement than
elaborate computations.  We briefly analyse the notion of the capacity
of a quantum communication channel.

\section{Elementary Quantum Gates, Algorithms and Implementation}

A quantum computer is a physical machine that can accept input states
which represent a coherent superposition of many different possible
inputs and subsequently evolve them into a corresponding superposition
of outputs. Computation, {\em i.e.\/} a sequence of unitary
transformations, affects simultaneously each element of the
superposition, generating a massive parallel data processing capability albeit
within one piece of quantum hardware~\cite{DD85}.  This way quantum
computers can efficiently solve some problems which are believed to be
intractable on any classical computer~\cite{DJ92,PWS94}. Apart from
changing the complexity classes, the quantum theory of computation
reveals the fundamental connections between the laws of physics and
the nature of computation and mathematics~\cite{DDBook}.

For the purpose of this paper a quantum computer will be viewed as a
quantum network (or a family of quantum networks) composed of quantum
logic gates; each gate performing an elementary unitary operation on
one, two or more two--state quantum systems called {\em
qubits\/}~\cite{DD89}. Each qubit represents an elementary unit of
information; it has a chosen ``computational'' basis
$\{|0\rangle,|1\rangle\}$ corresponding to the classical bit values
$0$ and $1$.  Boolean operations which map sequences of 0's and 1's
into another sequences of 0's and 1's are defined with respect to this
computational basis.

Any unitary operation is reversible and that is why quantum networks
effecting elementary arithmetic operations such as addition,
multiplication and exponentiation cannot be directly deduced from
their classical Boolean counterparts (classical logic gates such as
{\small\sf AND} or {\small\sf OR} are clearly irreversible: reading
$1$ at the output of the {\small \sf OR} gate does not provide enough
information to determine the input which could be either $(0,1)$ or
$(1,0)$ or $(1,1)$). Quantum arithmetic must be built from reversible
logical components. It has been shown that reversible networks (a
prerequisite for quantum computation) require some additional memory
for storing intermediate results~\cite{CHB89}.  Hence the art of
building quantum networks is often reduced to minimising this
auxiliary memory or to optimising the trade--off between the auxiliary
memory and a number of computational steps required to complete a
given operation in a reversible way.

We show three elementary gates that are used in construction of more
complicated quantum networks: The Not gate (which is obviously
reversible), Controlled-Not gate and the Toffoli gate.
\begin{figure}[h]
\begin{center}
\epsfxsize=9cm 
\epsfbox{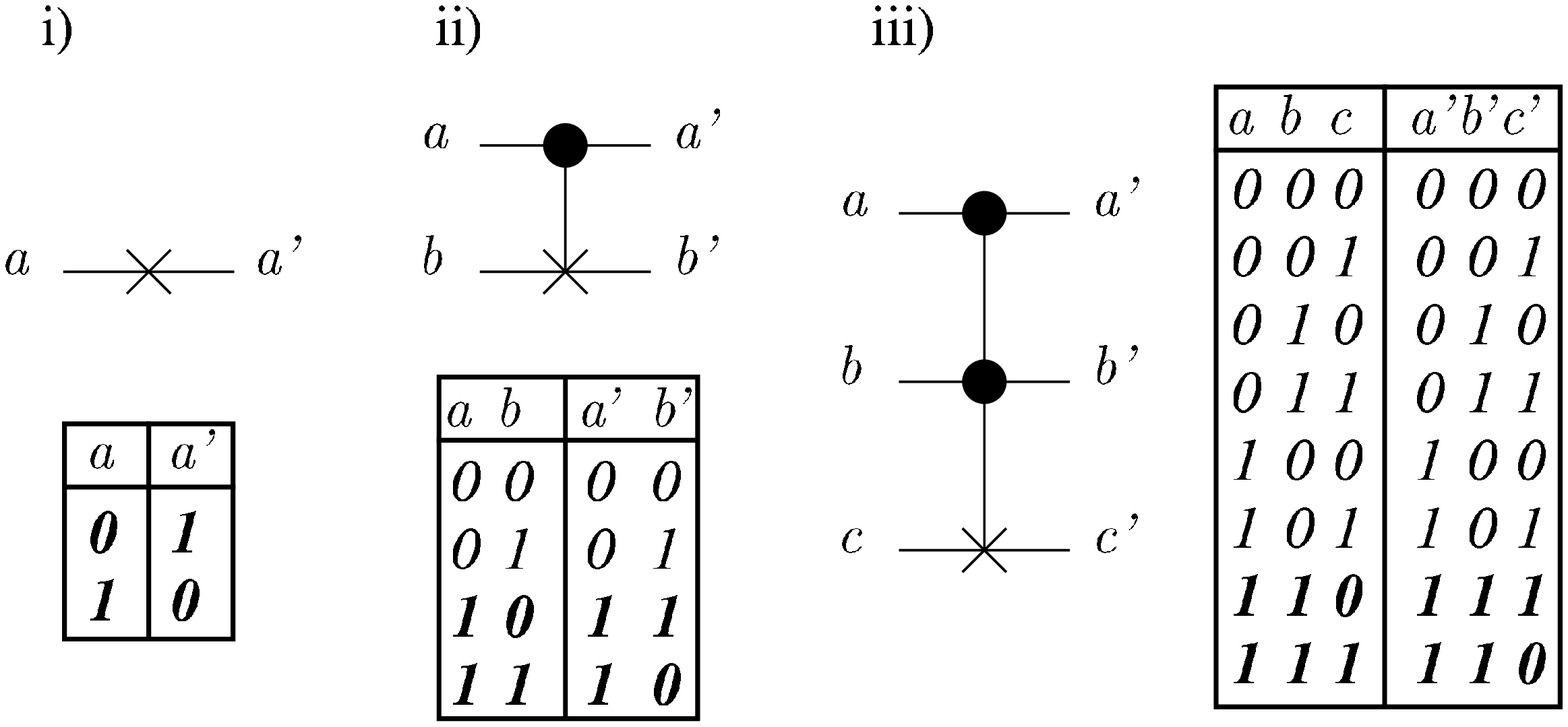}
\end{center}
\caption{\small{Elementary gates: (i) Not gate, (ii) Controlled Not 
and (iii) Toffoli gate}}
\end{figure}

By {\em basic} quantum gates we mean any set of quantum gates which
can perform any desired quantum computation. A universal quantum gate
is the one whose combination can be used to simulate any other quantum
gate. A number of quantum algorithms have been developed (see
elsewhere in this volume) from Deutsch's oracle algorithm to Shor's
factorisation algorithm and Grover's search algorithm.  All may be
realized in principle using networks made up from one-bit rotations
and CNOT gates. The Shor algorithm for factorisation uses Euclid's
method and periodicity to find the factors of the given number N.
This requires addition, multiplication and exponentiation networks and
Fourier transformation \cite{Vedral0}.  The Grover search algorithm
solves the problem of finding a special entry within a database of
length $N$.  Classically we need $N/2$ tries, but a quantum computer
can find the entry in $\sqrt{N}$ tries \cite{Grover}.

We will not provided an exhaustive review of all possible
implementations of quantum logic gates here.  Many have been proposed,
from coupled quantum dots, NMR spins, laser cooled ions coupled
through their centre of mass motion, to cavity qed in which atomic
superpositions become entangled with quantised single mode cavity
fields.  Quantum gate operation has been demonstrated experimentally
for some of these, and we will concentrate in what follows on the
special case of the linear ion trap gate.  This involves cooling ions
to the lowest quantised state of motion within a trapping potential
and then entangling internal and motional degrees of freedom of the
trapped ions.  Meekhof et al (1996) have shown how a number of
nonclassical motional states of a ${\rm Be}^+$ ion may be realized;
their experiments reveal they are limited to some extent by dephasing
decoherence.  Nevertheless, the same trap has been utilised to realize
a CNOT gate \cite{Monroe1}. In what follows, we discuss the problem of
decoherence in such a realization.

\section{Decoherence Problems}

The ion trap CNOT gate involves cooling ions to their lowest
vibrational state within the trap potential.  Then single-photon (or
two-photon Raman transitions) can excite internal electronic
transitions within the ion; a suitable choice of detuning can
simultaneously create (or annihilate) vibrational quanta.  Meekhof et
al (1996) showed in particular how Fock states of motion can be
realized by a clever choice of laser pulses and detunings.  Were there
to be {\bf no} sources of decoherence, the trapped ion dynamics should
reflect the Jaynes-Cummings interaction of internal and vibrational
degrees of freedom \cite{Shore}. For a Fock state, this would be a
pure sinusoidal Rabi oscillation.  What was observed \cite{Meekhof1}
was a damped Rabi oscillation of the form
\begin{eqnarray}
P_\downarrow \left( t \right)
= \frac{1}{2} \left \{ 1
+\sum_n{p_n \cos{\left(B_n t\right) } 
{\rm e}^{-A_n t}} \right \},
\end{eqnarray}
where $P_\downarrow \left( t \right)$ is the probability of being in
$\left \vert \downarrow \right \rangle$ internal ion state, $p_n$ is
the initial vibrational quantum number probability distribution, $B_n$
is the coherent effective Rabi frequency and $A_n$ is a
phenomenologically-introduced decoherence rate.  These are
substantially larger than expected, and are observed to be n-dependent
as $A_n=\gamma_0 \left( 1+n \right)^{0.7}$.  Possible sources of this
decoherence include imperfect phase correlation for the field driving
the Raman excitations and heating of the motional states.  In what
follows, we show how such decoherence affects the qubit-vibrational
Jaynes-Cummings dynamics.

In the Lamb-Dicke limit of closely confined ion motion, the effective
Hamiltonian for the trapped ion experiment \cite{Meekhof1} in the
interaction picture is given by the Jaynes-Cummings Hamiltonian,
\begin{eqnarray} 
H_{eff}^{I}=\hbar g \left( a S_+ + a^\dagger S_- \right),
\label{eqn:hinteff}
\end{eqnarray}
where $a$, $a^\dagger$ are boson operators for the motional states
($\left \vert n \right \rangle_m$), and $S_{+}$, $S_{-}$ are spin
operations for the two relevant internal atomic levels ($\left \vert
\downarrow \right \rangle_a$ and $\left \vert \uparrow \right
\rangle_a$.  The Jaynes-Cummings Hamiltonian (\ref{eqn:hinteff}) is
the origin of the characteristic quantum dynamics of the system.  In
this section, we introduce phenomenologically new sources of
decoherence in the interaction picture, which destroy this
characteristic Jaynes-Cummings dynamics without energy relaxation
\cite{Murao1}. We formulate the effects of decoherence using a master
equation describing the coupling of the internal and vibrational
states to a quantum reservoir.  In the high temperature limit of the
reservoir, within Markovian approximation, the master equation
coincides with that for stochastic white noise.  The advantage of
using this quantum reservoir is that it not only describes quantum
noise, but also provides a microscopic understanding of decoherence.

The effects of an environment coupled to the Jaynes-Cummings system
are treated by coupling a quantum reservoir, which consists of an
infinite number of bosons in a canonical distribution at temperature
$T$ for each mode.  The choice of the coupling between the system
operators and the reservoir operators determines the effect of the
reservoir.  If we choose the system operators that do not change the
bosonic quantum number when they operate on the dressed states, the
resulting master equation describe relaxation {\bf within the dressed
states} indicated by the bosonic quantum number $n$, but not energy
relaxation between states with different $n$. The operators $S_z$,
$a^\dagger a$ are obviously of this type, as these operators do not even
change the motional states $\left \vert n \right \rangle_m$ as well as
the dressed state label $n$.  The operator $a S_+ + a^\dagger S_-$
changes the motional state, but this operator does not change the
dressed state indication $n$, so $a S_+ + a^\dagger S_-$ is of this
type, too.

We consider in the following two possible alternatives for
system-reservoir coupling as potential candidates for the source of
``decoherence without energy relaxation'':
\begin{eqnarray}
H_{sr}&=&\hbar \left( a S_+ +a^\dagger S_-\right) 
\sum_l{g_{l}^\prime \left( B_l^\dagger+B_l\right)}, 
\label{eqn:coupleSb}\\
H_{sr}^\prime&=&\hbar a^\dagger a \sum_l{g_{l}^\prime \left(
B_l^\dagger+B_l\right)},
\label{eqn:couplebb}
\end{eqnarray}
where $\omega_l$ is the $l$th reservoir frequency, and $B_l^\dagger$
and $B_l$ are the creation and annihilation operators of the reservoir
bosons.  The coupling (\ref{eqn:coupleSb}) describes imperfect dipole
transitions between the level $\vert 0 \rangle_a $ (the intermediate
state for the Raman transitions) and the level $ \vert j \rangle_a$
($j=\downarrow,\uparrow$) due to fluctuations of the driving laser
intensity. The coupling (\ref{eqn:couplebb}) describes fluctuations of
the trap potential.

Then the master equation for the reduced system operator in the
interaction picture $\rho^I \left(t\right)$ due to the
system-reservoir coupling is obtained using a time convolution-less
(TCL) formalism \cite{Shibata1} and the rotating wave approximation on
the master equation \cite{Murao2}
\begin{eqnarray}
    \frac {\partial}{\partial t}\rho^I \left(t\right)=
    \frac {1}{i\hbar}
    \left[ {H_{eff}^I,\rho^I \left( t \right)}\right]
     +{\it \Gamma} \rho^I \left( t \right) 
\label{eqn:mastereq1}
\end{eqnarray}
with the damping term $\it{\Gamma} \rho^I \left( t \right)$ given by
\cite{Murao2}
\begin{eqnarray}
\it{\Gamma} \rho^I \left( t \right) 
&=& \sum_l{{g_{l}^\prime}^2 \int_0^t{dt' \biggl\{\left( \langle
B_l^\dagger\left(t'\right)B_l\rangle_B+
\langle B_l\left(t'\right)B_l^\dagger\rangle_B \right) \biggr. }}
\nonumber\\
&\times& \left(\left[{C_s}\left(-t'\right) 
\rho^I\left( t \right),C_s^\dagger \right]+
\left[C_s^\dagger \left(-t'\right) \rho^I\left( t \right),
C_s \right]
\right) \nonumber\\
&+&\left( \langle B_l^\dagger\left(-t'\right)B_l \rangle_B+
\langle B_l\left(-t'\right) B_l^\dagger \rangle_B \right) 
\nonumber\\
&\times& \biggl. \left(\left[C_s,
\rho^I\left( t \right) C_s^\dagger \left(-t'\right)\right]+
\left[C_s^\dagger,
\rho^I\left(t\right) C_s \left(-t'\right) \right]
\right)\biggr\} 
\label{eqn:dampint}
\end{eqnarray}
where $C_s$ represents the system operators $a S^+$ and $a^\dagger a$,
which couple to the reservoir. Time evolution of the system
operators are determined by (\ref{eqn:hinteff}).

The master equation (\ref{eqn:mastereq1}) can be solved by expanding
all system operators in terms of the dressed states, which are
eigenstates of the effective Hamiltonian (\ref{eqn:hinteff}), under
certain reservoir conditions \cite{Murao2}.  We take the continuum
limit of the reservoir modes. We also require the time scale of the
reservoir variables to be much shorter than the system variables so we
can take the Markovian limit. If we assume an initial condition of a
product state $\vert \downarrow \rangle_a \langle \downarrow \vert
\otimes \sum_n p_n \vert n \rangle_m \langle n \vert$, the population
of the lower atomic state, given by 
\begin{eqnarray}
P_\downarrow \left( t \right)
= \frac{1}{2} \left \{ 1
+\sum_n{p_n \cos{\left(B_n t\right) } 
{\rm e}^{-A_n t}} \right \},
\label{eqn:pdownthoery}
\end{eqnarray}
is obtained from the analytical solution of an off-diagonal element of
the density matrix in the dressed state basis
$\rho_{12}^{nn}\left(t\right) ={\rm e}^{\left(-A_n \pm i B_n \right)t}
\rho_{12}^{nn}\left(0 \right)$.  The damping rate $A_n$ is
\begin{eqnarray}
A_n&=& \left(n+1 \right) \kappa \left( n \right)
\left \{ \hat{n} \left( n \right)+1/2 \right \} \equiv A_n^{di},
\label{eqn:damprate}\\
A_n&=& \frac{1}{2} \kappa \left( n \right)
\left \{ \hat{n} \left( n \right)+1/2 \right \} \equiv A_n^{vi},
\label{eqn:damprate2}
\end{eqnarray}
for the imperfect dipole transition case (\ref{eqn:damprate}), and for
the fluctuation of the vibrational potential case
(\ref{eqn:damprate2}), where ${\bar n \left( n \right) }$ is the mean
reservoir boson number given by $\hat{n} \left( n \right) =\left( {\rm
e}^{2 \hbar g \sqrt{n+1}/k_B T} -1 \right)^{-1}$, and $\kappa \left(n
\right)$ is assumed to be given $\kappa \left(n \right) \approx \left
(2 \hbar g \sqrt{n+1} \right)^d$. The effects of the zero frequency
reservoir bosons are neglected.  The coherent part $B_n$ is given by
$B_n=\sqrt{4 g^2\left( n+1 \right)-A_n^2}$.

The results for the decoherence rates $A_n^{di}$ (\ref{eqn:damprate})
and $A_n^{vi}$ (\ref{eqn:damprate2}) show that decoherence originates
in the relaxation of density matrix elements that are diagonal in the
boson quantum number but off-diagonal in the spin quantum numbers in
the dressed state basis.  This relaxation is caused by the coupling to
reservoir bosons at frequency of $2 g \sqrt{n+1}$.  The effective
contribution of reservoir bosons at frequency of $2 g \sqrt{n+1}$ is a
key to understand the decoherence rate.  The Rabi frequency $g$ in the
experiment \cite{Meekhof1} is around 100 KHz, so reservoir bosons of
order $100$ KHz may be responsible for decoherence.  These reservoir
bosons have much low frequency than those responsible for spontaneous
emission of atomic states, which is in order of GHz, and also
population decay of motional states, which is of order 10 MHz.  This
low frequency nature of the reservoir boson suggests that the
reservoir has {\it high temperature} nature whereas in the optical
frequency regime, the corresponding reservoir is often approximated at
zero temperature.  Thus we can have the high temperature limit.  This
limit represents the classical noise where the reservoir operators
commute.  Introducing normalised values,
$\tilde{A}_n^{di}=A_n^{di}/g$, $\tilde{A}_n^{vi}=A_n^{vi}/g$,
$\tilde{\gamma}_0=\gamma_0/g$, $\tilde{\kappa}\left(n \right)=\kappa
\left(n \right)/g$, the normalised decoherence rates are
\begin{eqnarray}
\tilde{A}^{di}&=&\tilde{\gamma}_0 \left( n+1 \right)^{(d+1)/2},\\
\tilde{A}^{vi}&=&\tilde{\gamma}_0 \left( n+1\right)^{(d-1)/2}.
\end{eqnarray}
To get an exponent of $0.7$ for the factor $\left (n+1 \right)$
suggested by the experiment \cite{Meekhof1}, we need $d \approx 0.4$
for the imperfect dipole transition case and $d \approx 2.4$ for the
case of fluctuations of the vibrational potential.
\begin{figure}[h]
  \begin{center}
    \leavevmode
    \epsfxsize=6cm
    \epsfbox{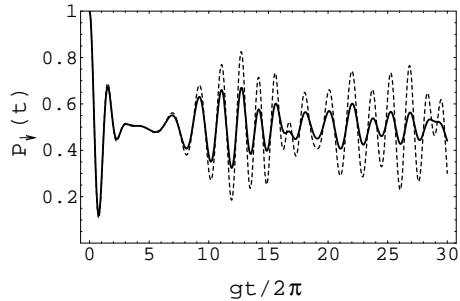}
  \end{center}
\caption{\small{The population of the lower atomic state $P_\downarrow \left
( t \right)$ with the initial state being the product of $\left \vert
\downarrow \right \rangle_a$ for the atomic state and a {\bf coherent}
state $\left \vert 3.0 \right \rangle_m$ for the motional state.  The
dashed line is for no decoherence and the solid line is for the case
of imperfect dipole transition with the coefficients $d=0.4$ and
${\tilde \gamma}_0=0.127$, which corresponds to the experiment of
Meekhof et al (1996).}}
\label{fig:fig}
\end{figure}

The sources of decoherence so far considered derive from instrumental
imperfections which are likely to improve in the future.  If we
imagine they can be entirely overcome, only fundamental sources of
decoherence such as spontaneous emission would remain.  We now examine
the consequences for quantum computation of this kind of decoherence
\cite{Plenio1,Plenio2}.  Spontaneous decay would terminate the
coherent superposition necessary for quantum computation.

An {\em elementary time step} (a coherent gate operation) takes the
time ${\tau}_{{\bf el}}$ and factorisation of an $L$ bit number
requires of the order of $\epsilon L^3$ elementary time steps where
$\epsilon$ is of order 400.  This results in a total computation time
$T$ of
\begin{equation}
	T \sim \epsilon\tau_{el} L^3\; .
	\label{1} 
\end{equation}
The {\bf decoherence time} of a single qubit is ${\tau}_{{\bf dec}}$ 
and the decoherence time for $5L+2$ (this number is required for 
factorisation) qubits is
\begin{equation}
	 \tau_{dec} = \frac{\tau_{qb}}{5L}.
	\label{2} 
\end{equation}
To prevent spontaneous emission during the computation we need
$\tau_{qb} \gg 5\epsilon\tau_{el} L^4$.  However, the larger the
decoherence time ${\tau}_{{\bf qb}}$ the longer is the elementary time
step ${\tau}_{{\bf el}}$!  \cite{Plenio1}

If we use a two level system as a qubit, then the coherent gate
operation is determined by the coherent Rabi frequency $\Omega_{12}$
But the Rabi frequency $\Omega_{12}$ and the spontaneous emission
decay rate $\Gamma_{22}$ are {\bf not independent}. We have
\begin{equation}
	\frac{\Omega_{12}^2}{\Gamma_{22}} = 
	\frac{6\pi c^3 \epsilon_0}{\hbar\omega_{12}^3} E^2,
	\label{5}
\end{equation}
where $E$ is the electric field strength of the laser
\cite{Plenio1,Plenio2}.  An upper limit for $E$ is the tunnelling
ionization field strength which for hydrogen has the value $E\cong
5.8\cdot 10^{11} V/m$.
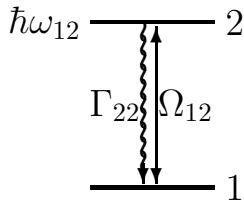
\begin{figure}[h]
\begin{center}
\setlength{\unitlength}{0.55mm}
\begin{picture}(80,40)
\thicklines
\put(50,0){\line(1,0){30}}
\put(50,0.3){\line(1,0){30}}
\put(50,40){\line(1,0){30}}
\put(50,40.3){\line(1,0){30}}
\put(66,3){\vector(0,1){36}}
\put(66,37){\vector(0,-1){36}}
\normalsize
\thicklines
\multiput(63,38)(0,-3.5){10}{\makebox(0,0)[c]{$\wr$}}
\multiput(62.8,38)(0,-3.5){10}{\makebox(0,0)[c]{$\wr$}}
\put(62.9,6){\vector(0,-1){5}}
\large
\put(56.,20){\makebox(0,0)[c]{$\Gamma_{22}$}}
\put(73,20){\makebox(0,0)[c]{$\Omega_{12}$}}
\put(39,40){\makebox(0,0)[c]{$\hbar\omega_{12}$}}
\put(85,0){\makebox(0,0)[c]{$1$}}
\put(85,40){\makebox(0,0)[c]{$2$}}
\normalsize
\end{picture}
\end{center}
\caption{\small{A two level system storing a quantum bit}}
\end{figure}

In the implementation of a CNOT in an ion trap, the COM mode has to be excited and
deexcited twice. This requires a full $4\pi$ rotation with the Hamiltonian
\begin{equation}
	H = \frac{\eta}{\sqrt{5L}}\,\frac{\Omega_{12}}{2}
	\left[ |e\rangle\langle g| a + |g\rangle\langle e|a^{\dagger}\right]
	\label{6}
\end{equation}
where $\eta$ is the Lamb-Dicke parameter, and $a$, $a^\dagger$ are the
vibrational annihilation and creation operators.  One needs $\epsilon
L^3$ elementary steps $\tau_{el}$ so that
\begin{equation}
	T \approx 4\,\pi\frac{\sqrt{5L}}{\eta\Omega_{12}} \, \epsilon L^3.
	\label{7}	
\end{equation} 
To have no spontaneous emission during the calculation we require
\begin{equation}
	\frac{1}{5L\,\Gamma_{22}} = {\tau}_{{\bf dec}} {\bf\gg T} =
	\frac{4\,\pi\epsilon}{\Omega_{12}}
	\sqrt{\frac{5L^7}{\eta^2}}.
	\nonumber
\end{equation}
This leads to 
\begin{equation}
	\frac{1}{\Gamma_{22}} \gg \frac{2000\pi^2\epsilon^2}{\eta^2}
	\frac{\Gamma_{22}}{\Omega_{12}^2}\, L^9.
	\label{9}
\end{equation}
For the total computation time we obtain
\begin{equation}
	T \gg 400 \pi^2  \left(\frac{\epsilon}{\eta}\right)^2
	\frac{\Gamma_{22}}{\Omega_{12}^2}\,L^8.
	\label{8}
\end{equation}
Some values for $T$ assuming $\eta=1,\Omega^2/\Gamma=10^{16} s^{1/2}$
and $\epsilon=500$ are shown in the following table
\begin{center}
\begin{tabular}{|c||c|c|}\hline
 $L$   &      $T\gg$      &     $\Gamma\ll$ \\ \hline
$4$    &  $0.0064\,s$         &  $77\cdot 10^{-2}\,s^{-1}$  \\ \hline
$40$   &  $6.4\cdot 10^{5}\,s$   &  $77\cdot 10^{-11}s^{-1}$ \\ \hline
\end{tabular}
\end{center}
For example, to factorize the $23$ digit number
\begin{equation}
41141158551285430224619 = 34802904313 \cdot 1182118543363
\end{equation}
on a quantum computer one needs about
\begin{equation}
	1.4\cdot 10^{8}s \approx 3.6\,\mbox{years}.
\end{equation}
Mathematica does it in $25s$ on a workstation! We have shown elsewhere
how breakdown of the two-state model for the qubit imposes even more
stringent restrictions on quantum computation \cite{Plenio2}.

These considerations showed the need to use quantum error correction
methods to stabilize the system against noise. However, quantum error
correction methods are implemented as short quantum computations
themselves and suffer from errors. To avoid this problem the new idea
of fault tolerant quantum computation \cite{PWS94,Shor2} was invented.
The idea is to encode the qubits in such a way that the encoding does
not introduce more errors than previously were present. If the error
stays at the same level we then keep performing error correction until
the error has decreased in magnitude \cite{Shor2,DiVincenzo1,Plenio3}. 
The present state of the art requires $5-10$ qubits to encode a single
qubit against a single error. It is the iterative application ``in
depth" of the encoding that will enable us to reduce error to an
arbitrarily small level providing it is below a certain level to start
with. In other words we will be encoding the encoding bits.

 We have seen above how to estimate the accuracy threshold for quantum
computation with a simple argument and we have given elsewhere
\cite{Plenio2} the numbers that arise from more precise explicit
constructions of error correction schemes. We have seen that the
incoherent error rate per quantum gate should not be higher than
around $10^{-6}$. In a more detailed analysis \cite{Knill1} it was
shown that the execution of one quantum gate on an encoded qubit
requires of the order of $N=10^6$ operations which confirms the
qualitative estimates given by arguments of the kind given above.  We
will now see whether accuracies of that order can be achieved in a
linear ion trap realisation of the quantum computer using as qubits
Zeeman sublevels in the chosen ions. We emphasise that we take into
account only the spontaneous emission of the ions and assume all the
other errors have been eliminated.

We calculate the probability to suffer at least one spontaneous
emission during the implementation of $N$ quantum gates. This
probability has to be smaller than unity. We represent the qubit by
two Zeeman sublevels and use Raman pulses to transfer population
between the two states. For the time required to perform $N$ quantum
gates we find $T = N 8\pi\Delta_2/\Omega_{02}^2$ From that we obtain
the probability for a spontaneous emission from level $2$ as $p_2 =
8\Gamma_{22}N/\Delta_{2}$.  Again we have to take into account the
fact that the two-level approximation can break down. This leads to an
additional independent source of spontaneous emission via extraneous
levels. One finally obtains the probability to have a spontaneous
emission from an extraneous level
\begin{equation}
	p_3 = \frac{80\Gamma_{33}^2\pi^2N^2 L}
	{\Delta_{13}^2\beta\eta^2}
	\left(\frac{\omega_{12}}{\omega_{13}}\right)^3.
	\label{102}
\end{equation}
The total probability of a spontaneous emission is $p_{tot}=p_2+p_3$ 
and therefore the error rate per quantum gate is 
\begin{equation}
	r = \frac{p_{tot}}{N} = \sqrt{\frac{320L}{\beta}} 
	\frac{\pi\Gamma_{33}}{\Delta_{13}\eta}
	\left(\frac{\omega_{12}}{\omega_{13}}\right)^{3/2}.
	\label{103}
\end{equation}
We use the data for the ions given in \cite{Plenio2}.  If we assume
$\eta=1,\beta=1$ \cite{Knill1} $L=7$ and an optimistic $N=10^6$ we see
that even for Barium the probability for at least one emission is
almost unity. The explicit values are for Barium $r=0.44\, 10^{-6}$,
for Mercury $r=9.26\, 10^{-6}$ and for Calcium $r=2.03\, 10^{-6}$.
This means that unless the encoding procedures given in
\cite{Knill1,Aharonov1} can be improved substantially the accuracy
threshold for quantum computation will not be achievable. Some
progress in this direction has been made recently \cite{Steane3}. We
conclude that the ion-trap computer is at present incapable of very large
scale computations, so we next look at some simpler, but equally
fundamental and useful problems, which can be achieved using such
realizations.

\section{Generalisation of Entanglement Swapping}

\begin{figure}[h] 
\begin{center} 
\leavevmode 
\epsfxsize=5cm
\epsfbox{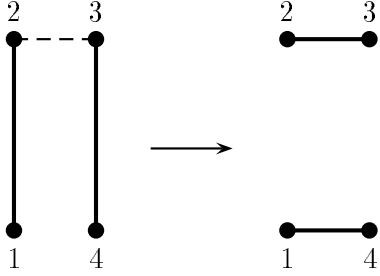}
\caption{\small{The swapping of entanglement between pairs of particles due
to a Bell state measurement on two of them is shown. The bold lines
connect particles in Bell states and the dashed lines connect
particles on which the Bell state measurement is made.}}
\end{center}
\label{e2} 
\end{figure}

There are many interesting manipulations of entanglement (though not
computations) that one can do with a limited number of qubits and as
such these are potentially testable applications. An interesting
scheme in this category is entanglement swapping. We first briefly
recapitulate the original version of this scheme \cite{swap}. Consider
an initial state of four particles 1, 2, 3 and 4 in which particles 1
and 2 are mutually entangled (in a Bell state), and particles 3 and 4
are mutually entangled (also in a Bell state). If one conducts a
measurement of the Bell operator on particles 2 and 3 (which projects
particles 2 and 3 to a Bell state), then the particles 1 and 4 are
instantaneously projected to one of the Bell states as well. Whereas
prior to the measurement, the Bell pairs were (1,2) and (3,4), after
the measurement the Bell pairs are (2,3) and (1,4). A pictorial way of
representing the above process is given in Fig.\ref{e2}. It is clear
that the most interesting aspect of this scheme is that particles 1
and 4, which do not share any common past, are entangled after the
swapping.

We have generalised the method of entanglement manipulation described
above to cases where a greater number of particles are involved
\cite{us}. But before that we need to introduce some notation and
terminology. In terms of a binary variable $u_i \in \{0,1\}$ and its
complement $ u_i^c $ (defined as $1-u_i$), one can write down any Bell
state (not normalised) of two particles $i$ and $j$ as
\begin{equation}
\label{b1}
     |\Psi(i,j)\rangle_{\pm} = |u_i, u_j \rangle \pm |u_i^c, u_j ^c\rangle ~.
\end{equation}
In the above it is understood that $|u_i\rangle$ and $|u_i^c\rangle$ are  two
orthogonal states of a two state system. Then N-particle generalisation of
Bell states will be states of the type

\begin{equation}
|\psi\rangle = \prod_{i=1}^{N} |u_i \rangle \pm \prod_{i=1}^{N} |u_i^c \rangle .
\end{equation}
For the N=2 they reduce to the Bell states and for N = 3 and 4 they
represent the GHZ states. For a general N we shall call them cat
states. We shall show that the original entanglement swapping scheme
can be generalised to the case of starting with cat states involving
any number of particles, doing local measurements by selecting any
number of particles from the different cat states and also ending up
with cat states involving any number of particles.  To see that
consider an initial state in which there are $N$ different sets of
entangled particles in cat states. Let each of these sets be labelled
by $m$ (where $m=1,2,..,N$), the $i$th particle of the $m$th set be
labelled by $i(m)$ and the total number of particles in the $m$th set
be $n_m$. Then the initial state can be represented by
\begin{equation}
\label{Nini}
        |\Psi\rangle = \prod_{m=1}^N |\Psi\rangle_m   ,
\end{equation}
in which each of the cat states $|\Psi\rangle_m $ is given by
\begin{equation}
\label{mini}
        |\Psi\rangle_m = \prod_{i=1}^{n_m} |u_{i(m)} \rangle \pm \prod_{i=1}^{n_m} |u_{i(m)}^c \rangle 
\end{equation}
where the symbols $u_{i(m)}$ stand for binary variables $\in \{0,1\}$
with $u_{i(m)}^c=1-u_{i(m)}$. Now imagine that the first $p_m$
particles from all the entangled sets are brought together (i.e a
total of $p=\sum_{m=1}^N p_m$ particles) and a joint measurement is
performed on all of them. Note that the set of all cat states of $p$
particles forms a complete orthonormal basis. Let the nature of the
measurement on the selected particles be such that it projects them to
this basis. Such a basis will be composed of states of the type
\begin{equation}
\label{e1fin}
 |\Psi(p)\rangle = \prod_{m=1}^N \prod_{i=1}^{p_m} |u_{i(m)} \rangle \pm \prod_{m=1}^N \prod_{i=1}^{p_m} |u_{i(m)}^c \rangle    .
\end{equation}
By simply operating with $|\Psi(p)\rangle \langle \Psi(p)|$ on
$|\Psi\rangle $ of Eq.(\ref{Nini}), we find that the rest of the
particles (i.e those not being measured) are projected to states of
the type
\begin{equation}
\label{e2fin}
 |\Psi(\sum_{m=1}^N n_m-p)\rangle = \prod_{m=1}^N \prod_{i=p_m+1}^{n_m} |u_{i(m)} \rangle \pm \prod_{m=1}^N \prod_{i=p_m+1}^{n_m} |u_{i(m)}^c \rangle    ,
\end{equation}
which represents a cat state of the rest of the particles. In a
schematic way the above process can be represented as
\begin{equation}
   \prod_{m=1}^N |E(n_m)\rangle \rightarrow |E(p)\rangle \otimes |E(\sum_{m=1}^N n_m-p)\rangle
\end{equation}
where $|E(n)\rangle$ denotes a $n$ particle cat state. As a specific
example, in Fig.5, we have shown the conversion of a collection of two
Bell states and a 3 particle GHZ state to a 3 particle GHZ state and a
4 particle GHZ state due to a projection of 3 of these particles to a
3 particle GHZ state.
\begin{figure}[h] 
\begin{center} 
\leavevmode 
\epsfxsize=5cm 
\epsfbox{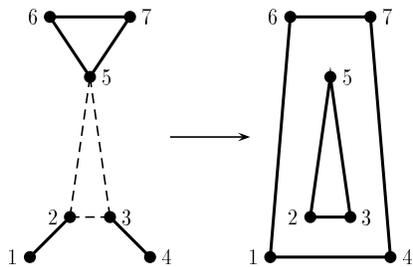}
\caption{\small{The conversion of two Bell states and a 3 particle GHZ state to a 3 particle GHZ state and
a 4 particle GHZ state due to a GHZ state projection on three
particles (one taken from each of the initially entangled sets) is shown. The bold lines connect mutually entangled particles and the dashed lines connect particles on which the GHZ state projection is made.}}
\end{center}
\label{k2} 
\end{figure}
 
As must be evident from Fig.5, there is a general "pencil and paper"
rule for finding out the result when our method of entanglement
manipulation is applied to a certain collection of cat states of
particles. One just has to connect the particles being measured to
frame a polygon and those not being measured to frame a complementary
polygon. These two polygons represent the two multiparticle cat states
obtained after the manipulation.

This scheme can be used for practical purposes such as constructing a
{\em quantum telephone exchange}, speeding up the distribution of
entangled particles between two parties and a sort of series
purification \cite{us}. We describe the first application in some
detail below.

\section{Quantum telephone exchange}

Suppose there are N users in a communication network.To begin with,
each user of the network needs to share entangled pairs of particles
(in a Bell state) with a central exchange. Consider Fig.6 : A, B, C
and D are users who share the Bell pairs (1,2), (3,4), (5,6) and (7,8)
respectively with a central exchange O. Now suppose that A, B and C
wish to share a GHZ triplet. Then a measurement which projects
particles 2, 3 and 5 to GHZ states will have to be performed at O.
Immediately, particles 1, 4 and 6 belonging to A, B and C respectively
will be reduced to a GHZ state. In a similar manner one can entangle
particles belonging to any N users of the network and create a N
particle cat state.
          
The main advantages of using this technique for establishing
entanglement over the simple generation of N particle entangled states
at a source and their subsequent distribution are as follows.

(A) Firstly, each user can at first purify a large number of partially
decohered Bell pairs shared with the central exchange to obtain a
smaller number of pure shared Bell pairs. These can then be used as
the starting point for the generation of any types of multiparticle
cat states of the particles possessed by the users. The problems of
decoherence during propagation of the particles can thus be avoided in
principle. Also the necessity of having to purify N-particle cat
states can be totally avoided. Purification of singlets followed by our
scheme will generate N-particle cats in their purest form.

(B) Secondly, our method allows a certain degree of freedom to
entangle particles belonging to any set of users only if the necessity
arises. It may not be known in advance exactly which set of users will
need to share a N particle cat state. To arrange for all possibilities
in an a priori fashion would require selecting all possible
combinations of users and distributing particles in multiparticle
entangled states among them. That is very uneconomical. On the other
hand, generating entangled N-tuplets at the time of need and supplying
them to the users who wish to communicate is definitely time
consuming.
\begin{figure}[h] 
\begin{center} 
\leavevmode 
\epsfxsize=5cm 
\epsfbox{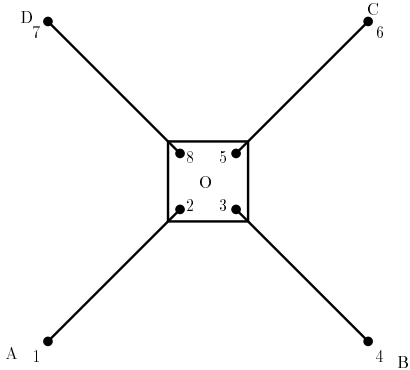}
\caption{\small{The configuration used for the distribution of
entanglement. Initially users A,B,C and D share Bell pairs with the
central exchange O. Subsequently, a local measurement at O is
sufficient to entangle particles belonging to any subset of users
chosen from A, B, C and D.}}
\end{center}
\label{s1} 
\end{figure}

It is pertinent to compare our scheme with the Biham-Huttner-Mor
cryptographic network with exchanges \cite{mor}. There are two main
differences: Firstly, they used a time reversed EPR scheme for setting
up the connections and had quantum memories to protect their
states. We use a multiparticle generalisation of entanglement
swapping. Secondly, their prime focus was to connect any pair of users
of a N-user network on request, while our main focus is to establish
multiparticle entangled states of the particles possessed by the users. 
Of course, for completeness, we must highlight some uses of
distributed multiparticle entanglement. An application that readily
comes to mind is multiparty cryptographic conferencing. We have found
out another interesting application. When $N+1$ users in a network
share one particle each from an $N+1$ particle cat state, then one of
these users can read messages sent by all the others through a single
measurement. This is a multiparticle generalisation of the superdense
coding scheme \cite{wies}. We have been able to show that though our
scheme uses far less number of particles, the rate at which a receiver
receives information in this scheme is the same as the rate at which
he would receive information if he was separately doing superdense
coded communication with each of the users \cite{us}.

\section{Quantum Communication}
Having demostrated how entanglement may be manipulated, we next turn to a
discussion of how it may be used to improve communication channel capacities. But first we need to quantify how much entanglement we posses within a given state.
We have recently shown how to construct a whole class of measures of
entanglement \cite{Vedral,Vedral2}, and also imposed conditions that
any candidate for such a measure has to satisfy\cite{Vedral}.  In
short, we consider the disentangled states which form a convex subset
of the set of all quantum states. Entanglement is then defined as a
distance (not necessarily in the mathematical sense) from a given
state to this subset of disentangled states.  An attractive feature of
our measure is that it is independent of the number of systems and
their dimensionality, and is therefore completely general,
\cite{Vedral,Vedral2}.  It should be noted that in much the same way
we can calculate the amount of classical correlations in a state. One
would then define another subset, namely that of all product states
which do not contain any classical correlations. Given a disentangled
state one would then look for the closest uncorrelated state. The
distance could be interpreted as a measure of classical correlations.

\begin{description}
\item E1. $E({\sigma})=0$ iff ${\sigma}$ is separable.
\item E2. Local unitary operations leave $E({\sigma})$ 
invariant, i.e.
$E({\sigma})=E(U_A\otimes U_B \sigma
U_A^{\dagger}\otimes U_B^{\dagger})$.
\item E3. The expected entanglement cannot 
increase under 
LGM+CC+PS given by $\sum  V^{\dagger}_i V_i = {{1 \hspace{-0.3ex} \rule{0.1ex}{1.52ex}\rule[-.01ex]{0.3ex}{0.1ex}}}$, i.e. 
\begin{equation}
\sum tr(\sigma_i) \,\, E( \sigma_i/tr(\sigma_i)) 
\le E(\sigma)\;\; , 
\end{equation}
where $\sigma_i = V_i \sigma V^{\dagger}_i$.
\item E4. $E(\sigma)$ is continuous.
\item E5. $E(\sigma)$ reduces to the von Neumann entropy for pure states.
\item E6. Additivity of $E(\sigma)$: $E(\sigma_1\otimes\sigma_2) = 
E(\sigma_1) + E(\sigma_2)$.
\end{description}
The only choice that we have found so far satisfying the above is 
\begin{equation}
E(\sigma) := \min_{\rho\in {\cal D}} S(\sigma ||\rho) \; ,
\end{equation}
where $S(\sigma ||\rho)= \mbox{tr} (\sigma \ln \sigma -
\sigma\ln\rho)$ is the quantum relative entropy. We call this measure
the {\em relative entropy of entanglement}.

What is interesting is that this quantity in addition 
represents an upper bound to any 
purification procedure (see e.g. \cite{puri}). 
Namely, if Alice and Bob start with 
an ensemble of $N$ entangled qubits in a state $\sigma$,
then the maximum number of singlets, $M$, distillable by local
operations is governed by the formula:
\begin{equation}
N E(\sigma) \ge M \ln 2
\end{equation}
This being so, we can easily see that $E(\sigma)$ is directly related
to the quantum capacity of a quantum communication channel. In a
quantum communication protocol Alice receives a quantum system in an
unknown state which she then wishes to transmit to Bob as reliably as
possible through a noisy quantum channel. They might use any quantum
resource including entanglement to achieve this.  For example, Alice
might create a maximally entangled pair, and send one of the particles
to Bob through the noisy channel.  Once they share a number of
partially entangled pairs they can purify them to singlets and then
use teleportation protocol for perfect transmission. In this case, the
rate at which Alice can transmit quantum information (i.e. the channel
capacity) will depend on how efficiently they can purify and that in
turn depends on the entanglement of the shared imperfect pairs. In
this case the capacity would be equal to $E(\sigma)$. It remains to be
seen whether this is the most efficient way of quantum transmission,
and until then the question of quantifying the quantum channel
capacity remains unclear \cite{Lloyd,Schumacher}.

\section{Conclusions}
We have studied the impact of spontaneous emission on the practical
applicability of quantum computation in linear ion traps and
especially the possibility of using a quantum computer to factorize
large numbers. We conclude that with present technology such a
factorisation will not be possible even if we employ sophisticated
methods of quantum error correction. We have shown that the numbers
that can be factorised will be restricted to almost trivial sizes. We
then investigated the minimal error rate per quantum gate and compared
it to recently established accuracy thresholds that would, in
principle, allow arbitrarily complicated quantum computations. We find
that the presently known thresholds cannot be achieved because of
spontaneous emission alone. Other sources of error would lead to even
stronger limitations. We conclude that new physical ideas are
therefore necessary if the goal of practically useful quantum
computation is to be reached. For this reason we have turned to applications of
which require only small-scale quantum systems. 

\section{Acknowledgements} 
This work was supported by a European Community Network, the UK
Engineering and Physical Sciences Research Council, by a Feodor-Lynen
grant of the Alexander von Humboldt Foundation, by the Japan Society
for the Promotion of Science, and by the Knight Trust.

\end{document}